\documentclass[aps,prb,superscriptaddress,twocolumn,epsf,float]{revtex4}
\usepackage{graphicx}
\usepackage{bm}
\usepackage{setspace}
\begin{document}
\title{Current induced torques and interfacial spin-orbit coupling:
Semiclassical Modeling}
\author{Paul M. Haney}
\affiliation{Center for Nanoscale Science and Technology, National
Institute of Standards and Technology, Gaithersburg, Maryland
20899-6202, USA }
\author{Hyun-Woo Lee} 
\affiliation{PCTP and Department of Physics, Pohang University of Science and Technology, Kyungbuk 790-784, Korea}
\author{Kyung-Jin Lee}
\affiliation{Korea University, Department of Material Science \&
  Engineerin, Seoul 136701, South Korea}
\affiliation{Korea Institute of Science and Technology, Seoul 136-791, Korea}
\affiliation{Center for Nanoscale Science and Technology, National
Institute of Standards and Technology, Gaithersburg, Maryland
20899-6202, USA }
\affiliation{Univeristy of Maryland, Maryland Nanocenter, College Pk, MD 20742 USA}
\author{Aur\'{e}lien Manchon}
\affiliation{Core Labs, King Abdullah University of Science and Technology (KAUST), Thuwal 23955-6900, Saudi Arabia}
\author{M. D. Stiles}
\affiliation{Center for Nanoscale Science and Technology, National
Institute of Standards and Technology, Gaithersburg, Maryland
20899-6202, USA }

\begin{abstract}
In bilayer nanowires consisting of a ferromagnetic layer and a non-magnetic
layer with strong spin-orbit coupling, currents create torques on the
magnetization beyond those found in simple ferromagnetic nanowires.  The
resulting magnetic dynamics appear to require torques that can be
separated into two terms, damping-like and field-like.  The
damping-like torque is typically derived from models describing the
bulk spin Hall effect and the spin transfer torque, and the field-like
torque is typically derived from a Rashba model describing interfacial
spin-orbit coupling.  We derive a model based on the Boltzmann
equation that unifies these approaches.  We also consider an
approximation to the Boltzmann equation, the drift-diffusion model,
that qualitatively reproduces the behavior, but quantitatively fails
to reproduce the results.  We show that the Boltzmann equation with
physically reasonable parameters can match the torques for any
particular sample, but in some cases, it fails to describe the
experimentally observed thickness dependences.
\end{abstract}

 \maketitle

\section{Introduction}
\label{sec:intro}

Spintronic applications like spin-transfer-torque magnetic random
access memory (STT-MRAM) or magnetic domain wall-based devices require
advances in materials to reach their full potential.  The goal of
improving these materials has led to the study of bilayers consisting
of ferromagnetic layers and non-magnetic layers with strong spin orbit
coupling.  Recent measurements on such systems have demonstrated
efficient switching of magnetic tunnel junctions,\cite{Liu:2012} like
those used in STT-MRAM, and efficient current-driven domain wall
motion.\cite{Miron:2011a}

There are a number of physical processes\cite{vanderBijl:2012} in
these systems that contribute to the magnetization dynamics as
described by the Landau-Lifshitz-Gilbert equation.  These include the
typical micromagnetic contributions, like interatomic exchange,
magnetostatic interactions, magnetocrystalline anisotropy, and
damping, as well as the adiabatic and non-adiabatic spin transfer
torques\cite{Berger:1978,Berger:1984,Zhang:2004,Tatara:2004,Thiaville:2005,Tserkovnyak:2008,Tatara:2008,Shibata:2011}
that are typically added to account for the coupling between the
magnetization and the electrical current flowing through it.  In the
bilayers of interest here, there are additional contributions that
have received extensive attention.  These arise from the spin-orbit
coupling in the non-magnetic layer and from the enhanced spin-orbit
coupling at the interfaces between layers.

These additional contributions have been modeled in terms of two
different pictures.  One picture\cite{Ando:2008} assumes that the
layers are thick and 
the two layers have their bulk properties. A
current flowing through the non-magnetic layer with strong spin-orbit
coupling generates a spin current perpendicular to the interface (the
spin Hall
effect.\cite{Hirsch:1999,Zhang:2000,Sinova:2004,Murakami:2003})  When
this spin current impinges on the interface, 
there is a spin transfer
torque\cite{Slonczewski:1996,Berger:1996,Stiles:2006,Ralph:2007} on 
the magnetization of the magnetic 
layer.  The details of the torque in this picture are determined by
the bulk spin Hall angle in the material with strong spin-orbit
coupling and the mixing conductance.  The other
picture\cite{Obata:2008,Manchon:2008,Manchon:2009,MatosAbiague:2009}
assumes two 
dimensional transport that can be described by a Rashba
model, similar to those used to describe spin-orbit
coupling in two-dimensional electron gases.\cite{Bychkov:1984} The
Rashba model gives direct coupling between the magnetization and the
flowing current.  Both models give qualitatively similar results, that
is torques along the ${\bf M}\times ({\bf j}\times \hat{\bf z})$ and
${\bf M}\times[{\bf M}\times ({\bf j}\times \hat{\bf z})]$ directions,
where ${\bf M}$ is the magnetization, ${\bf j} $ is the in-plane current
density and the interface normal is in the $\hat{\bf z}$ direction.
We refer to the first torque as a field-like torque because it has the
same form as precessional torque around an effective field in the
$-{\bf j}\times \hat{\bf z}$ direction.  The second torque has the
same form as a damping torque toward a field in that same direction
and we refer to it as a damping-like torque.\cite{torque}

Both models have strengths and weaknesses.  The Rashba model treats
the strong spin-orbit coupling at the interfaces between the materials
but treats the transport as two-dimensional.  The layer thicknesses
are usually comparable to mean free paths and spin diffusion lengths
requiring a three dimensional description of the transport.  On the
other hand, the spin-Hall-effect spin-transfer-torque model treats the
three dimensional aspect of the transport, but ignores any
contributions from the modification of the spin-orbit coupling near
the interface.  The non-magnetic layer and the magnetic layer affect
the electronic structure of each other close to the interface and the
interaction can significantly change the spin-orbit coupling there.
In particular, the proximity to the ferromagnet can induce a moment in
the material with strong spin-orbit coupling giving a thin layer where
the magnetism and the spin-orbit coupling coexist.\cite{RashbaDistorted}

Attempts to develop predictive models face the complication that the
experimental structures deviate significantly
from the ideal structures treated theoretically.  Experimental
indications\cite{Lavrijsen:2012,parkin:unpub} that interfaces of Co
grown on Pt have 
different properties than interfaces of Pt grown on Co argue strongly
that the details of the interface structure are both nontrivial and
important.  Unfortunately, the interfaces are not well enough
characterized to know what types of disorder might be present.  There
may be significant interdiffusion at the interfaces because, for example,
Pt alloys with Co in the bulk.  There is also significant lattice
mismatch between the materials.  This mismatch could promote thickness
fluctuations and dislocation formation.  Without measurements of
atomic scale structure of the experimental samples, it is impossible
to know how important such defects are to the behavior of the system.

Motivated by the uncertainty in the details of the experimental
structures and the goal of incorporating the strengths of existing
models, we develop simple semiclassical models for these systems.  One
approach is based on the Boltzmann equation and the other on the
drift-diffusion equation.  Both capture the essential physics of the
models that have been used so far and provide a test for whether a
model based on bulk properties and enhanced spin-orbit coupling at the
interface can account for the experimental behavior.  We find that these
models are general enough to reproduce the torques measured in any
single sample for
reasonable values of the parameters, but not all samples
with a single parameterization.  In Sec.~\ref{sec:exp} we
summarize the experimental evidence for the different interpretations.
We also summarize the micromagnetic simulations that provide the basis
for these interpretations.  
We briefly present the semiclassical transport models in Sec.~\ref{sec:be} and
then apply them in Sec.~\ref{sec:shstt} to the model for a bulk spin
Hall current in the non-magnetic layer leading to a spin transfer
torque at the interface with the ferromagnet.  In
Sec.~\ref{sec:rashba} we add in the interfacial spin orbit coupling
and show that this
captures the important physics that is included in the Rashba-model
approaches.  

\section{Experimental Results and Theoretical Implications}
\label{sec:exp}

Recent experiments on multi-layer structures report evidence for
current-induced torques due to spin-orbit coupling. These torques are
reported to be large enough to modify the magnetization dynamics and
may be utilized to facilitate spintronic applications.  Various
experiments report conflicting results for the size and direction of
the torque.  Many of these values are inferred from measurements of
the resulting dynamics in conjunction with simulations.  This section
aims to summarize the evidence from experiments, in conjunction with
simulations, for different forms of the torque. 

\begin{figure}
\includegraphics[width=0.9\columnwidth]{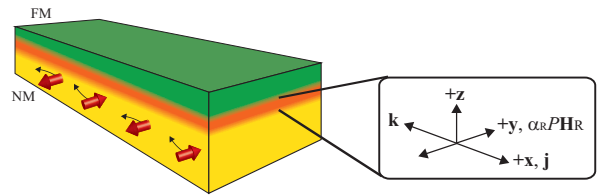}
\caption{(color online) Schematic of a bilayer structure that
consists of a ferromagnetic layer and a non-magnetic metal layer
with strong spin-orbit coupling. The spin Hall effect in the
non-magnetic layer bends electron trajectories and injects
electrons with proper spin direction into the adjacent magnetic
layer, thereby generating the damping like torque. In this
illustration, the spin Hall angle for the non-magnetic layer is
assumed to be positive. The figure also highlights the
interface region between the two layers, where the Rashba
spin-orbit coupling may have sizable magnitude.}
\label{fig:bilayer}
\end{figure}

The recent growth of interest in bilayer systems began with a series
of experiments by Liu {\it et
  al}.\cite{Liu:2011,Liu:2011b,Liu:arXiv,Liu:2012} and Miron {\it et
  al}.\cite{Miron:2010,Miron:2011a,Miron:2011b}  Both sets of
experiments treat systems of a substrate, non-magnetic layer,
ferromagnetic layer, and capping layer, see Fig.~\ref{fig:bilayer},
although the details of each differ from experiment to experiment.
The authors of the first set of experiments interpret their results in
terms of a dominant damping like torque that they attribute to the
spin Hall effect and as subsequent spin transfer torque.  On the other
hand, the authors of the second set of experiments interpret their
results in terms of an important field like torque that they attribute
to interfacial spin orbit coupling.

In the case of the spin Hall
effect, electron trajectories are preferentially scattered in different
directions 
depending on their spin directions. For instance, in a
non-magnetic material with positive spin Hall angle, electrons are
scattered more strongly into directions such that $({\bf v}_{\rm
  i}\times{\bf v}_{\rm f}) 
\cdot {\bf S}$ is positive, where ${\bf v}_{\rm i(f)}$ is the
electron velocity before (after) the scattering and ${\bf S}$ is the
electron spin direction. Thus in the bilayer system in
Fig.~\ref{fig:bilayer}, the spin Hall effect in the non-magnetic
layer injects electrons with spin along $+\hat{\bf y}$ direction
into a magnetic layer if ${\bf j}$ is along $+\hat{\bf x}$
direction and the spin Hall angle is positive. As it does for
perpendicular flow of electrons in multilayer systems, this spin
current causes spin transfer
torques,\cite{Slonczewski:1996,Berger:1996,Stiles:2006,Ralph:2007} the
injected electrons giving rise to a spin torque along the
direction $-{\bf M}\times ({\bf M}\times {\bf m})$, where ${\bf
m}$ is the direction of the magnetic moment carried by injected
electrons and points along $-\hat{\bf y}$ direction since ${\bf
m}$ and ${\bf S}$ are anti-parallel. This torque amounts to the
damping-like torque $-{\bf M}\times[{\bf M}\times ({\bf j}\times
\hat{\bf z})]$ arising from the spin Hall effect.

Liu {\it et al}. examine various magnetization dynamics caused by
the spin Hall effect. In Pt/Py bilayer,\cite{Liu:2011} they use
the spin-torque ferromagnetic resonance technique to show that the
spin Hall effect is strong enough to cause magnetic
precession. Through the resonance line shape analysis, they 
quantify the magnitude of the spin-Hall-effect-induced damping
like torque and the torque due to the Oersted field.
They estimate the spin Hall angle of Pt to be about
+0.076, which is about two orders of magnitude larger than the
corresponding value in $n$-doped GaAs.\cite{Kato:2004,Engel:2005}
This demonstrates that the spin Hall effect can be a realistic
tool to enhance spin torque efficiency. In their subsequent
work\cite{Liu:2012} for the Ta/Co$_{40}$Fe$_{40}$B$_{20}$
bilayer, they demonstrate that the spin-Hall-induced damping-like
torque can switch the magnetization in a reliable and efficient
way, facilitating the development of magnetic memory and
nonvolatile spin logic technologies. From three different
measurements (spin-torque ferromagnetic resonance, current-dependent
anomalous Hall effect, threshold current density for
magnetization switching), they estimate the spin Hall angle of Ta
to be $-0.12$ to $-0.15$. This angle is of opposite sign compared
to Pt but most notably, larger by about a factor of two. In their still
later work,\cite{Liu:2011b} they demonstrate that the
magnetization switching can be achieved by the spin Hall effect in
Pt(2~nm)/Co(0.6~nm)/AlO$_x$ systems as well, although the spin
Hall angle in Pt is smaller.  Most recently, they have determined an
even larger spin Hall angle in W.\cite{Pai:arxiv2012}

On the other hand, Miron {\it et
al}.\cite{Miron:2010,Miron:2011a,Miron:2011b} explore effects of
the Rashba spin-orbit coupling with focus on the system Pt(3
nm)/Co(0.6~nm)/AlO$_x$(2~nm). The layer structure of the film
breaks inversion symmetry and gives rise to perpendicular magnetic
anisotropy strong enough that the ground state for the
magnetization is out of plane.  In these experiments, the
nominally symmetric control system Pt(3~nm)/Co(0.6~nm)/Pt(3~nm),
shows much weaker effects, leading to the inference that the
broken inversion symmetry for the layer plays a crucial role.

Miron {\it et al} interpret their results in terms of a large
field-like torque as expected from the Rashba model\cite{Bychkov:1984}
for interfacial spin-orbit coupling.
Calculations\cite{Obata:2008,Manchon:2008,Manchon:2009,MatosAbiague:2009}
predict an effective field\cite{comment-factor2}
\begin{equation}
\mu_0 {\bf H}_{\rm R}\approx \frac{\alpha_{\rm R}}{2\mu_{\rm
B}M_{\rm s}}P\left(\hat{\bf z}\times {\bf j}\right),
\label{Eq:Rashba-field}
\end{equation}
when a system is subject to the Rashba spin-orbit coupling of the
form $\alpha_{\rm R}({\bf k}\times \hat{\bf z})\cdot \bm{\sigma}$.
Here $M_{\rm s}$ is the saturation magnetization, $P$ is the spin
polarization, and $\mu_{\rm B}$ is the Bohr magneton. The
effective field ${\bf H}_{\rm R}$ generates the field-like torque
$-\gamma {\bf M}\times {\bf H}_{\rm R}\propto -{\bf M}\times (
\hat{\bf z} \times {\bf j})$, where $\gamma$ is the gyromagnetic
ratio.

In Ref.~\onlinecite{Miron:2010}, 
the authors measure the reversal of
the perpendicular magnetization.  They find that transverse in-plane magnetic
fields enhance the nucleation and that this enhancement is modified by
currents flowing through the film.  In-plane currents ${\bf j}$
flowing through the system enhance (suppress) the nucleation of
reversed magnetic domains when the direction $\hat{\bf z}\times {\bf
  j}$ is parallel (anti-parallel) to externally applied in-plane
magnetic fields.  In the nominally symmetric control sample, the
effect of in-plane current is much weaker.  The authors interpret the
effect of in-plane currents as evidence for the current-induced
effective field along $\hat{\bf z}\times {\bf j}$, as predicted
theoretically for systems with Rashba-like spin orbit coupling.
Experimentally,\cite{Miron:2010} the magnitude of the effective field
is proportional to ${\bf j}$ with a proportionality constant $(1.0\pm
0.1)\times 10^{-12}$ T/(A$\cdot$ m$^{-2}$).  This value implies
$\alpha_{\rm R}\approx$ 0.2 eVnm,\cite{comment-factor2-2} which is
comparable to $\alpha_{\rm R}=0.3$ eVnm reported for Bi/Ag(111)
surface\cite{Ast:2007} and comparable to a recent first principles
calculation result for a Pt/Co bilayer.\cite{Park:2013}  

Further evidence for a field-like torque is found in
measurements\cite{Miron:2011a} of current-driven domain wall (DW)
motion in the same system. As the driving current density goes up,
the domain wall velocity $v_{\rm DW}$ increases to
speeds up to $\approx 400$~m/s, more than three times
faster than previously measured current-driven domain wall
velocities.\cite{Hayashi:2007}  The authors claim that this
velocity is twice as large as the rate $v_{\rm s}=|{\bf
j}|(Pg\mu_{\rm B})/(2eM_{\rm S})$ of the spin angular momentum
transfer, where $g(\approx 2)$ is the gyromagnetic ratio. Even up
to such high domain wall speeds, the domain walls apparently did
not undergo structural instability (Walker
breakdown.\cite{Beach:2005,Schryer:1974}) According to
conventional theories\cite{Zhang:2004,Thiaville:2005} of
current-driven domain wall motion, domain wall velocities above
$v_{\rm s}$ are possible below the breakdown current
density $j_{\rm W}$.  However, as the ratio $v_{\rm DW}/v_{\rm s}$
increases, the breakdown current density decreases such that
conventional theories cannot explain the velocities measured in this
experiment.\cite{Miron:2011a}

\begin{figure}
\includegraphics[width=0.9\columnwidth]{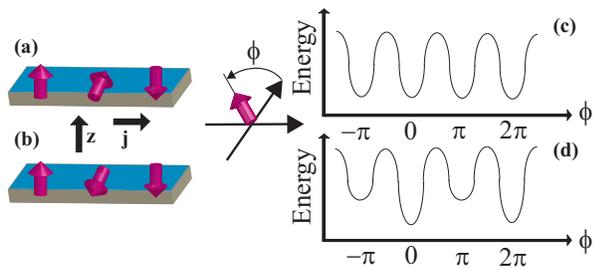}
\caption{(color online) Two possible structures of the Bloch
domain wall. In (a), the magnetization at the center of the
domain wall points along $\hat{\bf z}\times {\bf j}$  and in (b), it
points along $-\hat{\bf z}\times {\bf j}$.  Panel (c) shows
the domain wall energy as a function of the domain wall tilting
angle $\phi$ without the effective field ${\bf
H}_{\rm R}$.  Panel (d) shows the energy with the effective field
assuming that $\alpha_{\rm R}P$ is 
positive and ${\bf H}_{\rm R}$ prefers $\phi=0$.}
\label{fig:DW-structures}
\end{figure}

As a possible explanation, the authors suggest\cite{Miron:2011a}
that the current-induced effective field ${\bf H}_{\rm R}$
could increase the breakdown current density.  Consider the two low energy
structures [Figs.~\ref{fig:DW-structures}(a) and (b)] of the Bloch
domain wall.  These walls differ from each other
because they have opposite magnetization directions at the domain
wall center.  Since the two directions are parallel and
anti-parallel to ${\bf H}_{\rm R}$, the effective field either
stabilizes or destabilizes the corresponding Bloch domain wall
structures [Fig.~\ref{fig:DW-structures}(d)].  
Recalling that the conventional spin
torques produced by the current tend to
shift\cite{Thiaville:2005} $\phi$ away from the low energy values
($\phi=0$ and $\pi$), the deeper energy valley implies
an enhanced threshold current density to escape the valley. Since the
Walker breakdown occurs when $\phi$ cannot remain stationary, the
Walker breakdown threshold current density becomes the larger of
the two, which is larger than the Walker
breakdown threshold value without ${\bf H}_{\rm R}$. 

Another interesting feature of the experiment\cite{Miron:2011a}
is that contrary to
previous
experiments,\cite{Yamaguchi:2004,Yamanouchi:2004,Hayashi:2007} which
report domain wall motion along the electron flow direction,
Ref.~\onlinecite{Miron:2011a} 
reports domain wall motion {\it
against} the electron flow. Within the scope of the
conventional theories,\cite{Zhang:2004,Thiaville:2005} the
reversed domain wall motion implies negative $P\beta$, where
$P$ is the polarization of the current and $\beta$ is the
dimensionless coefficient of the non-adiabatic spin transfer
torque.  All of the qualitative feature of the
experiment\cite{Miron:2011a} would be
explained if there were an ${\bf H}_{\rm R}$ that gives Walker breakdown
threshold  
enhancement, negative $P\beta$ for reversed motion, and
$|\beta/\alpha|>1$ for
velocity enhancement ($\alpha$ is the Gilbert damping constant).

\begin{figure}
\includegraphics[width=0.9\columnwidth]{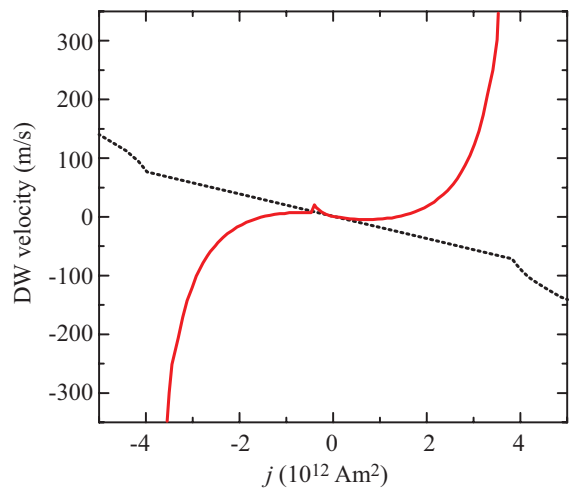}
\caption{(color online) Terminal domain wall velocity $v_{\rm DW}$
as a
  function of the current density ${\bf j} = j \hat{\bf x}$. The dotted
  line gives the prediction of the Landau-Lifshitz-Gilbert equation
  when the current generates only the conventional (adiabatic and
  nonadiabatic) spin transfer torques and the solid line the
  prediction when the current also generates field-like and
  damping-like torques.  For this calculation,
  we consider a nanostrip 
  of length $\times$ width $\times$ thickness = 2000~nm $\times$ 200~nm
  $\times$ 5~nm, with perpendicular magnetic anisotropy and the
  materials parameters as follows: $\gamma /(2 \pi)=28$~GHz/T,
  $M_{\rm s}
  = 1 \times 10^6$~A/m, $A_{\rm ex} = 1.3 \times 10^{-11}$~J/m, $P = 0.7$,
  $K_{\rm u} = 1.5 \times 10^6$~J/m$^3$, $\alpha = 0.5$, $\beta = 0.25$, and
  $\alpha_{\rm R} = 0.7 \times 10^{-10} $~eVm.  For the conventional
  calculation (dotted curve), the
  non-linearities are due to Walker breakdown. In the solid curve, the
  small blip at a small negative current is caused by the chirality
  switching of domain wall due to the spin-orbit-related field-like
  torque.}
\label{fig:DW-velocity}
\end{figure}

Further measurements\cite{Miron:2011b} on the same Pt(3
nm)/Co(0.6~nm)/AlO$_x$(1.6~nm) system reveal a problem in this
simple theoretical picture.  Based on measurements of bipolar switching
in a tilted field,
the authors conclude that the switching is
due the additional presence of a damping-like torque along
 ${\bf M}\times [{\bf M}\times (\hat{\bf
z}\times {\bf j})]$.
The damping-like torque is either parallel or
anti-parallel to the nonadiabatic STT
at the domain wall center and thus modifies the domain wall
velocity.\cite{Kim:2012b,Seo:2012} A recent theoretical
study\cite{Kim:2012b} demonstrates that high $v_{\rm DW}$ {\it
against} electron flow direction is possible
(Fig.~\ref{fig:DW-velocity}) even when both $P$ and $\beta$ are
positive and $\beta/\alpha$ is smaller than 1, if both the
field-like torque $-{\bf M}\times (\hat{\bf z}\times {\bf j})$ and
the damping-like torque ${\bf M}\times [{\bf M}\times (\hat{\bf
z}\times{\bf j})]$ are sufficiently large.

Miron {\it et al.}\cite{Miron:2011b} estimate that the spin Hall
contribution\cite{Liu:2011b} to the damping-like torque is not
strong enough to explain the bipolar switching and conclude that
the damping-like torque arises mainly from the Rashba spin-orbit
coupling, based on the observation that the efficiency of the
bi-polar switching increases with the magnetic anisotropy of the
cobalt layer and the oxidation of the aluminum layer.
Calculations\cite{Wang:2012,Kim:2012b,Pesin:2012,vanderBijl:2012}
confirm that the Rashba spin-orbit coupling can give rise to the
damping-like torque. Note however that the spin Hall contribution
and the Rashba spin-orbit coupling contribution to the
damping-like torque have exactly the same structure, making the
distinction difficult. In this regard, recent micromagnetic
calculations\cite{Kim:2012b} show that the spin Hall effect and
the Rashba spin-orbit coupling contributions give
qualitatively different domain wall motion if there is no field-like torque as
might be expected if the spin Hall effect is the dominant source of
the effect. Without the field-like torque,
the domain wall motion at high current densities is along the electron flow
direction (if $\beta P>0$) whereas with the strong field-like torque,
the domain wall motion 
against the electron flow is possible even for $\beta P >0$ (solid
line in Fig.~\ref{fig:DW-velocity}).

In contrast to Miron {\it et al}.'s
interpretation,\cite{Miron:2011b} Liu {\it et
al.}\cite{Liu:2011b} conclude that the damping-like torque in
Pt(2.0~nm)/Co(0.6~nm)/Al(1.6~nm) arises mainly from the spin Hall
effect in Pt and the Rashba spin-orbit coupling contribution is
negligible. As evidence for the latter
conclusion, they report that the ratio between the current-induced
effective field along $\hat{\bf z}\times {\bf j}$ and
the in-plane current density is more than 75 times smaller than
the corresponding ratio in 
Ref.~\onlinecite{Miron:2010}. 
A third
experiment\cite{Pi:2010} on a similar system
Pt(3.0~nm)/Co(0.6~nm)/AlO$_x$(1.8~nm) found that this ratio is about
3.4 
times
smaller than the ratio reported in 
Ref.~\onlinecite{Miron:2010}.

Experiments\cite{Suzuki:2011} that measure the field-like
torque on
Ta(1.0~nm)/Co$_{40}$Fe$_{40}$B$_{20}$(1.0~nm)/MgO(2.0~nm) find that
the 
ratio of the effective field to the current is about 23~\% of the
corresponding ratio in Pt/Co/AlO$_x$ 
as reported in 
Ref.~\onlinecite{Miron:2010}. 
Interestingly, the
current-induced effective field in the Ta/CoFeB/MgO system is
along $-\hat{\bf z}\times {\bf j}$ and thus opposite to the
corresponding direction in the Pt/Co/AlO$_x$ system if the Ta and
MgO layers in the Ta/CoFeB/MgO system are matched with Pt and
AlO$_x$ layers in the Pt/Co/AlO$_x$ system, respectively. It was
also reported that the torque to current ratio decreases by more than
an order of 
magnitude when the thickness of the Co$_{40}$Fe$_{40}$B$_{20}$
layer in the Ta/CoFeB/MgO system increases slightly from 1.0~nm to
1.2~nm.

The thickness dependence was investigated
systematically\cite{Kim:2013} for the two wedge systems,
Ta($d_{\rm Ta}$)/Co$_{20}$Fe$_{60}$B$_{20}$(1~nm)/MgO(2~nm) and
Ta(1~nm)/Co$_{20}$Fe$_{60}$B$_{20}$($t_{\rm CoFeB}$)/MgO(2~nm).
When the Ta layer thickness $d_{\rm Ta}$ changes by 1~nm, the
effective field along the $\hat{\bf z}\times {\bf j}$ direction
changes its magnitude by nearly two orders of magnitude. For small
$d_{\rm Ta}\lesssim 0.6$~nm, the sign of the effective field is
opposite to that in the larger $d_{\rm Ta}$ regime and agrees with
the sign of the Pt/Co/AlO$_x$ system.\cite{Miron:2010} The
damping-like effective field along the $(\hat{\bf z}\times {\bf
j})\times {\bf M}$ direction is also sensitive to $d_{\rm Ta}$
with the sign change at $d_{\rm Ta}\approx$ 0.5~nm. The sign
change of the damping-like effective fields is interpreted as an
evidence of competition between Rashba spin-orbit coupling and
spin Hall effect.\cite{Liu:2012}

Measurements of current-driven domain wall motion in multi-layer
structures containing non-magnetic heavy metal layers and
ferromagnetic layers but without oxide
layers\cite{parkin:unpub,Haazen:2012} are reported. The motion is
also very sensitive to layer thicknesses.\cite{parkin:unpub} The
domain wall velocity can be up to almost 1~km/s in certain
multi-layer structures with the Pt layer thickness of about 1~nm
but changes very quickly as the Pt thickness varies.
Interestingly, very high domain wall speed ($\approx$ 1~km/s)
are observed only when the domain wall moves against the electron
flow, implying that the origin of the reversed domain wall motion
is probably correlated with the mechanism behind very high domain
wall speed. Several other
experiments\cite{Lee:2011,Lavrijsen:2012,Ono:unpub,Beach:unpub}
also report reversed domain wall motion in ultrathin multi-layer
systems containing Pt layers.

The analysis of domain wall motion described above considers four
current induced torques, the the adiabatic and non-adiabatic
spin-transfer torques and two torques, damping-like and field-like,
that depend on the layer structure of the device.  The first two
torques depend on the gradient of the magnetization but are determined
by bulk properties.  The last two are independent of the gradient of
the magnetization.  Other possibilities that depend on the gradient of
the magnetization and the layer structure are allowed by
symmetry\cite{vanderBijl:2012} and may be important for the dynamics.
In addition, recent calculations\cite{Thiaville:2012} suggest that a
current-independent torque due to the Dzyaloshinskii-Moriya
interaction\cite{DMInteraction} might provide an alternate mechanism
for stabilizing a moving domain wall above the nominal
Walker-breakdown field.  In this paper we only consider the
current-induced torques that are independent of the gradient of the
magnetization. 

\section{Semiclassical models}
\label{sec:be}

To explore possible mechanisms for the torques operative in these
systems, we develop semiclassical models that allow for easy
exploration of parameter space.  We use a Boltzmann equation approach
and the simpler drift-diffusion approach.  The Boltzmann
equation is better suited to describe in-plane transport but the
drift-diffusion approach is simpler and provides a useful language to
describe the physics.

The Boltzmann equation approach developed by Camley and
Barna\'{s}\cite{Camley:1989} is the simplest model that describes
current-in-the-plane (CIP) giant-magnetoresistance (GMR).  In the
Boltzmann equation, the variables are the distribution functions.  The
distribution function accounts for electrons moving in all directions
even though the total current only points in a single direction.  This
generality allows the approach to describe the flow of spins between
the layers even though the current flows in the plane of the
interfaces.

The drift-diffusion approach of Valet and Fert\cite{Valet:1993} is
based on integrating the distribution function in the Boltzmann
equation to derive transport equations that depend on the densities
and currents.  It has had wide success describing
current-perpendicular-to-the-plane (CPP) GMR, but does not describe
CIP GMR.  It fails because it does not describe the flow of spin
currents between the layers when the current flows in the plane of the
layers.  However, this limitation may be less important in the bilayer
systems of interest here.  In materials with strong spin-orbit
coupling, like Pt, spin currents do flow perpendicular to the charge
current because of the spin Hall effect\cite{SpinHall} so that the
drift-diffusion approach does qualitatively describe the physics.  However,
we compare the two approaches below, and show that the drift-diffusion
approach differs quantitatively from the Boltzmann equation for
similar reasons to its qualitative failure for CIP GMR.

Reference~\onlinecite{Xiao:2007} 
describes the matrix Boltzmann
equation we use in this paper.  It is a generalization of the model
used to describe CIP GMR,\cite{Camley:1989} and is based on a very
simplified model for the electronic
structure.  We treat all Fermi surfaces as spherical and as having the
same Fermi wavevector.  This approach ignores the details of the Fermi
surfaces, which are undoubtedly important for specific systems,
particularly when including spin-orbit coupling.  However, the
scattering mechanisms are both unknown and uncharacterized, so we feel
that it is appropriate to consider models in which scattering rates
and other physical processes are parameterized and the details of the
electronic structure are neglected for simplicity.
By performing appropriate integrals over the distribution function,
the Boltzmann equation can be transformed into a drift-diffusion
equation like that given in 
Ref.~\onlinecite{Brataas:2005}. 
The
parameterized processes in the Boltzmann equation then have a simple
connection to those in the drift-diffusion equation.

One such process that we include through parameterization is the
spin-dependent conductivity in the ferromagnetic layer.  We model the
spin dependence by using spin-dependent scattering rates.  Such
spin-dependent scattering is physically sensible as it is believed to
play a bigger role in the polarization of the current in these
materials than the electronic structure itself.  For example, in Co
and Ni, the current is expected to be dominated by majority carriers,
even though there are more minority carriers at the Fermi surface.  In
the drift diffusion model, the spin-dependent scattering leads to a
different conductivity for the majority electrons $\sigma^\uparrow$
than the minority electrons $\sigma^\downarrow$.  This difference is
parameterized in terms of the spin polarization of the current, defined through
$P=(\sigma^\uparrow - \sigma^\downarrow)/(\sigma^\uparrow +
\sigma^\downarrow)$.  The drift-diffusion transport equations in the
ferromagnet are
\begin{eqnarray}
  {\bf j}&=&\sigma \bm\nabla \mu - P \sigma \bm\nabla
  (\hat{\bf M} \cdot  \bm\mu^{\rm s}) ,
\label{eq:DD_FMc}
\\
  Q_{ij}&=& \frac{\hbar}{2e}\hat{M}_j P\sigma \nabla_i \mu -
  \frac{\hbar}{2e}\sigma \nabla_i
  \mu^{\rm s}_j ,
\label{eq:DD_FMs}
\end{eqnarray}
where ${\bf j}$ is the charge current density, and ${\bf Q}$ is the tensor
spin current density where the first index is the spatial component
and the second index the spin component. $\mu$ is the
electrochemical potential, such that negatively charged electrons
diffuse against the gradient giving an overall positive sign for the
first term in Eq.~(\ref{eq:DD_FMc}).  Similarly
$\bm{\mu}^{\rm s}$ is the spin chemical
potential, which is a vector along the direction of the spin
accumulation, and the unit vector $\hat{\bf M}$ is the direction of
the magnetization.  The minus sign in the second term of
Eq.~(\ref{eq:DD_FMc}) arises because 
majority electron spins are aligned opposite to the magnetization.
These two signs are typical of possible sources of confusion in this
subject matter.  They arise because with the charge on the electron is
negative and angular momenta and moments are in opposite directions.

The equality of the Fermi surfaces would also allow for perfect
transmission of electrons across the interface between the materials.
In the Boltzmann equation, we include spin-dependent reflection by the
addition of a spin-dependent sheet potential (delta function) at the
interface.  Choosing the strength of this delta function allows us to
tune the spin-dependent interface resistance to any arbitrary
value.\cite{Stiles:2000} In the drift-diffusion approach, the
spin-dependent reflection becomes a spin-dependent interface
resistance or conductance as used in the closely related circuit
theory.\cite{Brataas:2000}

While the overall structure of the Boltzmann equation approach is the
same as that published earlier, there are some modifications.  One
difference is the treatment of dephasing.  In a ferromagnet, spins on
different parts of the Fermi surface precess at different rates and
travel with different velocities.  These differences, combined with
scattering between different parts of the Fermi surface, cause the
precessing spins to rapidly become out of phase with each
other.\cite{Stiles:2002} 
In Ref.~\onlinecite{Xiao:2007},
the transverse spin accumulation and current are forced to zero in the
ferromagnet to account for this dephasing of the transverse spin
population.  Here, we allow for transverse spin accumulation in the
ferromagnet but build in rapid spin precession and explicitly account
for the processes that cause dephasing.  The simplified model of the
Fermi surfaces that we use can lead to underestimation of dephasing
processes.  We have tested this approximation by adding an explicit
dephasing term.  While such a term quantitatively changes the spin
accumulation in the ferromagnet, we find that it does not change the
calculated torques.

The same dephasing process is absent in a drift-diffusion model but
can be included by adding an explicit dephasing term.  In this
approximation, the steady-state continuity equations in the
ferromagnet are
\begin{eqnarray}
 \bm\nabla \cdot {\bf j} &=& 0
\\
 \nabla_i Q_{ij} &=&
 - \frac{1}{\tau_{\rm ex}} ({\bf s} \times \hat{\bf M})_j
 - \frac{1}{\tau_{\rm sf}} {s}_j
 \nonumber \\
  & &
 - \frac{1}{\tau_{\rm dp}} \left[ \hat{\bf M} \times
  ( {\bf s} \times \hat{\bf M} )\right]_j
\label{eq:spincont}
\end{eqnarray}
where the spin accumulation ${\bf s}$ is proportional to the spin
chemical potential ${\bf s}=\cal{N}_{\rm s}\bm{\mu}^{\rm s}$ and the
precession time $\tau_{\rm ex} = \hbar/\Delta$ is related to 
the exchange splitting  between the magnetization and the spin
accumulation.  
Repeated indices are summed over.  The first term on the right hand
side of Eq.~(\ref{eq:spincont}) is the precession in the exchange
field, the second term is the spin flip scattering that reduces all
components of the spin accumulation, and the last term is the
dephasing that reduces only the parts of the spin accumulation
transverse to the magnetization.  Setting the transverse spin
accumulation to zero, as done in earlier Boltzmann equation
calculations\cite{Xiao:2007} and in magnetoelectronic circuit
theory,\cite{Brataas:2000} is equivalent to taking the limit that the
dephasing time goes to zero.

Both the size of the spin Hall effect and its underlying mechanism are
controversial.  The theory for the spin Hall effect is related to that
for the anomalous Hall effect, a subject that has been controversial
for decades.\cite{ahreviews} Measurements of the spin Hall angle (the
ratio of spin Hall and charge conductivities) for various materials
span a range of values.  Part of the variation may result from the
sensitivity of the extraction of the spin Hall angle from experimental
data to other material parameters needed to model the
experiments.\cite{Liu:arXiv} Measurements\cite{Liu:2012} show that the
spin Hall effect in Ta is bigger and of the opposite sign of that in
Pt, in agreement with previous calculations.\cite{Tanaka:2008} The
agreement between these trends in theory and experiment argues for an
intrinsic origin of the effect.  However, the calculated spin Hall
conductivity for Pt appears to be approximately an order of magnitude
too small in comparison to the measured value.  Spin Hall angles of
approximately the right order of magnitude have been
computed\cite{Gradhand:2010} for the extrinsic contributions of
various impurities in Cu and Au.

With this uncertainty in the mechanism for the spin Hall effect, we
use the form of scattering appropriate for the extrinsic skew
scattering contribution for computational simplicity.  In the
Boltzmann equation, we include skew
scattering as described by Engel et al.\cite{Engel:2005} but
generalize their results to include scattering that leads to the
inverse spin Hall effect in addition to the scattering that gives rise
to the spin Hall effect.  These scattering terms connect the current
with a perpendicular spin current and vice versa.  Both our approach
and the earlier work\cite{Engel:2005} neglect the scattering processes
that couple spin currents to spin currents moving in other directions.
Such process contribute to spin relaxation, which we include as a
phenomenological spin flip scattering process.  Details of the
scattering and the
default parameters we use are given in Appendix~\ref{app:be_scatt}.

In the non-magnetic material, the explicit forms of the charge and
spin currents in the drift-diffusion approximation we use
are\cite{Brataas:2005}
\begin{eqnarray}
\bf{j} &=& {\sigma} \boldsymbol{\nabla} \mu
-{\sigma_{\rm SH}} (\boldsymbol{\nabla} \times
\boldsymbol{\mu}^{\rm s}),
\label{EQ:DD_NMc}
\\
Q_{ij} &=& -\frac{\hbar}{2e}\sigma {\nabla}_i \mu^{\rm s}_j
-\frac{\hbar}{2e} \sigma_{\rm SH} \epsilon_{ijk} {\nabla}_k
\mu .
\label{Eq:DD_NMs}
\end{eqnarray}
where $\sigma$ is the conductivity, $\sigma_{\rm SH}$ is the spin Hall
conductivity coupling the spin and charge currents to the charge and
spin potentials, and $\epsilon_{ijk}$ is the Levi-Civita symbol.  As
with the Boltzmann equation, we neglect a term corresponding to the
spin Hall effect coupling the spin current to the spin potential,
assuming that it is small.

In both models, the torque on the magnetization is given by the
torque between the magnetization and the spin accumulation
\begin{eqnarray}
  {\bf T}= 
  \frac{\gamma}{\tau_{\rm ex}M_{\rm s}} {\bf M} \times {\bf s} 
+ \frac{\gamma}{\tau_{\rm dp}M_{\rm s}} {\bf M} \times({\bf M} \times {\bf s} )
 ,
\label{eq:torque_density}
\end{eqnarray}'
where the gyromagnetic ratio $\gamma=g\mu_{\rm B}/\hbar$ converts from angular
momentum (spin density) to magnetization (so ${\bf T}$ is a term in the
Landau-Lifshitz-Gilbert equation\cite{torque}). 
The second term, which is not present in the Boltzmann
equation calculations, captures the torque due to the dephasing of the
electron spins as they precess in the exchange field.

If there is no coupling of angular
momentum into the lattice (spin-orbit coupling or spin-flip
scattering) it is straightforward to relate this torque,
Eq.~(\ref{eq:torque_density}), to the 
divergence of the spin current.
In the ferromagnet, where there is spin-flip scattering but no other
spin-orbit coupling, the corrections to torque being simply the
divergence of the spin current can be found from
Eq.~(\ref{eq:spincont}).
For the components perpendicular ($\perp$) to the magnetization, the
torque is
\begin{eqnarray}
{\bf T}&=&-\gamma\left(1-{\beta}\right)\left({\bm\nabla}\cdot{\bf
    Q}\right)_\perp
    +\gamma{\beta}{\hat{\bf M}}\times({\bm\nabla}\cdot{\bf
    Q}),\nonumber\\
&&\beta=\frac{\tau_{\rm ex}}{\tau_{\rm sf}}\frac{1}{1+\xi^2},\;
\xi=\frac{\tau_{\rm ex}}{\tau_{\rm dp}}+\frac{\tau_{\rm ex}}{\tau_{\rm sf}},
\label{eq:torqueJ} 
\end{eqnarray}
where $\left({\bm\nabla}\cdot{\bf Q}\right)_\perp=-{\hat{\bf
    M}}\times[{\hat{\bf M}}\times({\bm\nabla}\cdot{\bf 
    Q})]$ is the component of the divergence of the spin current that
is perpendicular to the magnetization.
Equation (\ref{eq:torqueJ}) indicates that the spin torque is in
general not simply given by the divergence of the spin current but
possesses an additional component ${\hat{\bf
    M}}\times{\bm\nabla}\cdot{\bf Q}$ arising from the presence of
spin relaxation.  For the parameter set used here, $\beta$ is
negligible.

An important difference with the previously published\cite{Xiao:2007}
formalism for the Boltzmann equation is the inclusion of spin-orbit
coupling at the interface.  This is done by including an additional
term in the interface potential
\begin{eqnarray}
  V({\bf r}) =  \frac{\hbar^2 k_{\rm F}}{m} \delta(z) \left[ u_0 + u_{\rm ex}
    \bm{\sigma} \cdot
    \hat{\bf m} + u_{\rm R} \bm{\sigma} \cdot ( \hat{\bf
      k}\times\hat{\bf z} ) \right] ,
\label{eq:intpot}
\end{eqnarray}
where the interface is in the $\hat{\bf z}$ direction at $z=0$, $u_0$
is the spin-independent part of the potential, $u_{\rm ex}$ is the
spin-dependent part of the potential that gives rise to spin-dependent
reflection, $u_{\rm R}$ is the Rashba contribution, with ${\bf k}$
being the wave vector of an electron scattering from the interface,
$k_{\rm F}$ is the Fermi wave vector, and $m$ is the electron mass.
This additional term
captures the form of spin-orbit coupling that is allowed for the
simple electronic structure assumed here.  For more realistic band
structures, the form would be much more complicated.  Unfortunately,
it is difficult to compare $u_{\rm R}$ with the $\alpha_{\rm R}$ used
in previous publications.  Doing so requires a procedure for reducing
the Hamiltonian for a three dimensional system to one for a
two-dimensional system.

In Eq.~(\ref{eq:intpot}), the last two terms can be combined to give a
wave vector dependent field direction $\hat{\bf u}({\bf k})$ and
strength $u_{\rm eff}({\bf k})$ such that $u_{\rm eff}({\bf k})
\hat{\bf u}({\bf k})= u_{\rm ex} \hat{\bf m} + u_{\rm R} \hat{\bf
  k}\times\hat{\bf z}$.  With respect to this direction, the 
majority and minority transmission and reflection amplitudes are
\begin{eqnarray}
  T &=& \frac{ik_{z}/k_{\rm F}}{ik_{z}/k_{\rm F}-(u_0 \pm u_{\rm eff})}
\\
  R &=& \frac{u_0 \pm u_{\rm eff}}{ik_{z}/k_{\rm F}-(u_0 \pm
    u_{\rm eff})}
\label{eq:tramps}
\end{eqnarray}
Since both the magnitude and phase of the transmission and reflection
amplitudes are different for the majority and minority spin
components, an electron spin oriented along some arbitrary direction
undergoes a finite rotation when transmitted or reflected.  A part of
the torque on the electron spin is due to the coupling to the exchange
field and a part due to the spin-orbit coupling (Rashba contribution).
The reaction torque 
on the magnetization can be computed from the exchange coupling
between the spin density at the interface and the exchange field
\begin{eqnarray}
  {\bf T} = \delta(z) \frac{\gamma}{M_{\rm s}}
\left( \frac{\hbar k_{\rm F} u_{\rm ex}}{m}\right)  {\bf s} \times
    {\bf M}
\label{eq:int_torque}
\end{eqnarray}
where the spin density ${\bf s}$ is calculated from the incoming wave
function and the transmission amplitudes.  Since the potential is
proportional to a delta function, the torque density diverges but is
finite when integrated over a finite thickness.

The treatment we use for the interfacial spin-orbit coupling in the
Boltzmann equation does not generalize easily to the drift-diffusion
equation because there are not any wave vector dependent quantities in
that model.  It may be possible to define a generalization of the
conductance matrix used in the magnetoelectronic circuit theory.  In
typical usage, the longitudinal spin components couple to each other and
the transverse spin components couple to each other, but the longitudinal
and transverse spin components do not couple.  With the Rashba interaction
included, all spin components would couple.

The Boltzmann equation based approach differs quite significantly from
the approach used in which the system is modeled with a two
dimensional Rashba model.  In our Boltzmann equation approach,
electron spins get kicked when they pass through the interface, but
they spend no time ``in'' the interface.  In the Rashba model, the
entire system is the interface so the electrons (and spins) are ``in''
the interface at all times.  In this case, there is a spin
accumulation that builds up in the interface.  This spin accumulation
gives rise to the strong field-like torque found in these models.  In
spite of this difference, we find that the both approaches give
qualitatively similar torques.

The remaining difference with the previously published formalism for
the Boltzmann equation is that the boundary conditions are different.
The previous version treated perpendicular transport and here we treat
in-plane transport.  Here, the outer boundaries are perfectly
reflecting, either diffusely, specularly, or somewhere in between.

\section{Spin Hall effect plus spin transfer torque}
\label{sec:shstt}

In this Section, we describe the behavior of the model in the absence
of spin-orbit coupling at the interface.  In this limit, the spin Hall
effect in the non-magnetic layer generates a spin Hall current that
propagates perpendicular to the interface with spins pointed
perpendicular to both the interface normal and the direction of the
current.  When this spin current hits the interface with the
ferromagnet, angular momentum is transfered from the flowing spins to
the magnetization as is typical for spin transfer torques in magnetic
multilayers.\cite{Slonczewski:1996,Berger:1996,Ralph:2007}

\begin{figure}
\includegraphics[width=0.8\columnwidth]{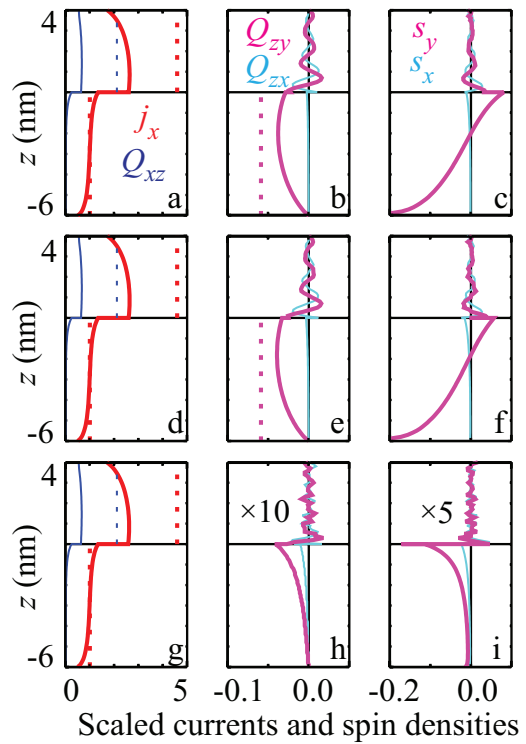}
\caption{(color online) Currents, spin currents, and spin
  accumulations.  The left panels (a, d, and g) show the current
  density (heavy lines) $j_x$, which is flowing in the plane of the
  sample and the spin current $Q_{xz}$ (lighter lines), flowing in the
  $x$-direction with spins aligned with the magnetization in the
  $z$-direction.  The dotted lines indicate the bulk values.  All
  currents and spin currents are dimensionless; currents are scaled by
  the bulk current in the non-magnet and spin currents are scaled by
  the bulk current in the non-magnet and an additional factor of
  $\hbar/2e$.  The 
  spin densities are scaled by the same factor two factors and $v_{\rm
    F}$.  The middle panels (b, e,
  and h) show the spin currents, $Q_{zy}$ (heavy lines) and $Q_{zx}$
  (lighter lines), flowing perpendicular to the layers ($z$-direction)
  with spins pointing perpendicular to the magnetization, i.e. the $x$- and
  $y$-directions.  The right panels (c, f, and i) show the
  accumulation of spin perpendicular to the magnetization, $s_y$
  (heavy lines) and $s_x$ (lighter lines).  The top panels (a, b, c)
  are for the case in which there is no interfacial spin orbit
  coupling, the bottom panels (g, h, i) for the case with
  interfacial spin-orbit coupling $u_{\rm R}=0.04$  and no spin Hall
  effect in the non-magnet, and the middle panels (d, e, f) for the
  case when both are present.  In panels (h) and (i), the spin
  accumulations have been scaled by the indicated factors.  } \label{fig:dist}
\end{figure}

This process is shown in Fig.~\ref{fig:dist} based on calculations
done with the Boltzmann equation described in Sec~\ref{sec:be}.
Parameter choices are
given in Table~\ref{tab:params} in Appendix~\ref{app:be_scatt}.  
Fig.~\ref{fig:dist} shows the currents, spin currents, and spin
densities for a 4~nm ferromagnetic layer coupled to a 6~nm
non-magnetic layer with the interface at $z=0$.
Panel
(a) shows the distribution through the thickness of the films of the
current flowing in the plane of the film (in the $x$-direction).  The
current is greater in the ferromagnetic layer because it has a higher
conductivity than the non-magnetic layer for this choice of
parameters.  The current is suppressed close to the outer boundaries
because we assume that the scattering from those interfaces is
completely diffuse.  In fact, the ferromagnetic layer is not thick
compared to the mean free paths, so the current is suppressed through
the thickness of the film.  The spin current with spins aligned with
the magnetization ($z$-direction) and moving in the plane also reduced
from the bulk value, in fact more so than the current, so the
polarization of the current is reduced from the bulk value.  At the
interface between the two materials, the current is enhanced in the
lower conductivity layer due to electrons entering from the higher
conductivity layer, and the current is reduced in the higher
conductivity layer.

This modification of the current near the interface is not captured by
a drift-diffusion model.  It is one source of the quantitative
disagreement between the models as discussed below.

Panel (b) of Fig.~\ref{fig:dist} shows the two components
of the spin current with spins aligned perpendicular to the
magnetization and moving perpendicular to the plane of the film.  In
the non-magnetic layer, this is due to the spin Hall effect.  The spin
current is zero at the lower boundary, which is both impenetrable and
has no spin-flip scattering.  It increases to close to its bulk value
at the interface between the non-magnet and the ferromagnet.  Inside
the non-magnetic layer, the spin current is a competition between the
spin Hall 
current and a diffusive spin current from the spin accumulation, seen
in Panel (c), that builds up due to the impenetrability of the outer
interface.  Panels (d-i) show calculations with interfacial spin-orbit
coupling included and are discussed in Sec.~\ref{sec:rashba}.

At and near the interface, this spin current is converted into a spin
transfer torque on the ferromagnet.  We write the interfacial torque
in the form 
\begin{eqnarray}
  {\bf T} &=& \delta(z) \frac{g\mu_{\rm B} j_0}{2e}
    \left[ \tau_{\rm d} \hat{\bf M}\times(\hat{\bf M}\times\hat{\bf y})
         + \tau_{\rm f} \hat{\bf M}\times\hat{\bf y} \right] ,
\label{eq:torqueform}
\end{eqnarray}
where $\delta(z)$ localizes the torque to the interface at $z=0$.
The dimensionless
coefficients $\tau_{\rm d}$ and $\tau_{\rm f}$ characterize the
``damping-like'' and ``field-like'' contributions respectively.  Other
terms are possible, as we discuss in 
Ref.~\onlinecite{RashbaDistorted}, 
but in
the three-dimensional transport calculations we find these other terms
to be negligible for the parameters we consider.

\begin{figure}
\includegraphics[width=0.8\columnwidth]{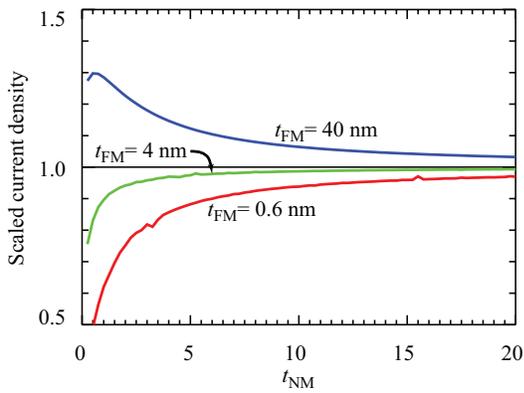}
\caption{(color online) Average current density in the non-magnetic
  layer.  For three thicknesses of the ferromagnetic layer, the
  average current density in the non-magnetic layer is shown scaled by
  the bulk value as a function of the thickness of the non-magnetic
  layer.}
\label{fig:current}
\end{figure}

The prefactor in Eq.(\ref{eq:torqueform}) is based on $j_0$, which is
the ``bulk'' current density in the non-magnetic layer, that is $j_0 =
\sigma_{\rm N} E$ where $E$ is the applied electric field.  This
choice seems to be that typically made in analyses of experiments even
though the total current is all that is directly measurable.  The rest
of the factors convert from current density to magnetization torque
density, $\dot{\bf M}$.  This choice makes sense in analyzing
experiments in terms of the spin Hall effect because the torque is
driven by the current density in the non-magnet.  However, for thin
films, there can be important corrections due to the outer boundaries
and the interface with the ferromagnet.  These corrections are shown
in Fig.~\ref{fig:current} for a variety of thicknesses for the two
layers.  The average current density is reduced by the diffuse
scattering assumed at the outer boundary of the layer, but is
increased by the (assumed) higher conductivity of the ferromagnetic
layer when that layer is thick enough.

For this model, with no
interfacial spin-orbit coupling, the spin transfer torque is
determined solely by the transverse spin current,\cite{Stiles:2002}
just outside the 
magnetic layer.  Since
neither the majority transmission probability is zero nor the minority
reflection probability is one, some of the transverse spin current is
reflected.  The reflected spin current is seen in the reduction of the
transverse spin current close to the interface.  Some of the
transverse spin current is absorbed right at the interface and some is
transmitted into the ferromagnet.  In the ferromagnet, spin components
transverse to the magnetization rapidly precess as they traverse the
layer, as seen in the oscillations in panel (b). Further, different
parts of the Fermi surface precess at different rates so they dephase
as the traverse the layer as can be seen by the decay of the transverse
spin current in the ferromagnet in panel (b).

\begin{figure}
\includegraphics[width=0.8\columnwidth]{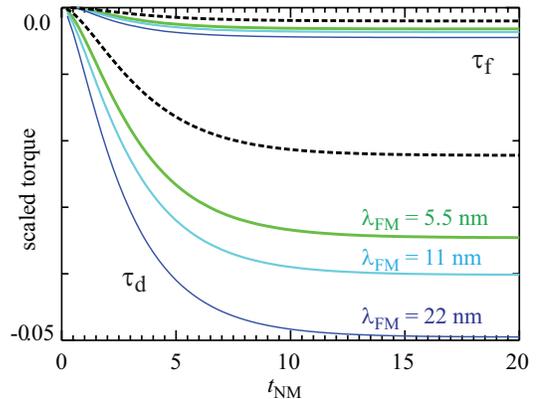}
\caption{(color online) Torques as a function of nonmagnetic layer
  thickness.  The solid curves are the full Boltzmann equation
  calculation and the dashed curves give the analytic approximation
  based on the drift-diffusion model and the circuit theory
  (Appendix~\ref{sec:analytic}).  The more negative curves 
  show the field-like torque $\tau_{\rm f}$ and those closer to zero
  show the damping-like torque $\tau_{\rm d}$.  For both torques, 
the Boltzmann results have been calculated for three different mean
free paths (labelled on the damping-like torques) in the ferromagnet.}
\label{fig:analytic}
\end{figure}

The dominant spin transfer torque arises from the absorption of the
incident transverse spin current either at the interface or in the
ferromagnet.  However, not all of the current is absorbed, and some is
rotated into the $x$-component of the spin current on reflection.  The
non-zero reflection reduces the damping like torque and the rotation
gives rise to a small field-like torque.  These torques are shown in
Fig.~\ref{fig:analytic} as a function of the thickness of the
non-magnetic layer.  

Figure~\ref{fig:analytic} shows the damping-like and field-like
torques calculated with both the Boltzmann equation
approach\cite{convergence} and the 
drift-diffusion approach.  The drift-diffusion approach gives an
analytic result, Eq.~(\ref{eq:analytic_td}) in
Appendix~\ref{sec:analytic}. That result is based on the
drift-diffusion model described in Sec.~\ref{sec:be} and
magnetoelectronic circuit theory\cite{Brataas:2000} for transport
across the interface.  Both approaches give the same behavior as a
function of the thickness of the non-magnetic layer.
Because the spin Hall current in the non-magnetic
layer is suppressed when the layer is thin, as seen in
Fig.~\ref{fig:dist}, the torque is reduced when the layer thickness is
less than a few spin diffusion lengths, which for this
set of parameters is $\ell_{\rm N}^{\rm sf}=2.5$~nm.  

For thick layers, the value saturates, but does not saturate to the
spin Hall angle $\theta_{\rm SH}$ as might be expected.
Eq.~(\ref{eq:analytic_td}) shows that for the drift-diffusion model,
the saturation value depends on the ratio, $\ell_{\rm N}^{\rm sf}{\rm
  Re}[g^{\uparrow\downarrow}]/\sigma_{\rm N}$.  When this ratio is
small, the saturation value is reduced from $\theta_{\rm SH}$ and the
Boltzmann calculation and the drift-diffusion calculation saturate to
different values.  When that ratio is large, the 
drift-diffusion and Boltzmann equation results agree.  However, a
large value of this ratio is not physically realistic for systems with
strong spin-orbit coupling.  The mixing conductance depends mainly on
the area of the Fermi surface in the non-magnet (as a reminder our
calculations assume the same Fermi surface for all materials), but
does so in the same way that the 
conductivity does (see the expression in Table~\ref{tab:params}, so it
is difficult to increase the ratio by 
changing the mixing conductance.  It is possible to decrease the
conductivity by increasing the non-spin-flip scattering, but this also
decreases the spin-diffusion length.  For the default parameters we
consider, see Table~\ref{tab:params}, the value of this ratio is about
0.3. 

Eq.~(\ref{eq:analytic_td}) also shows that the torque calculated with
the drift-diffusion approach is independent of the details of the
ferromagnetic layer, depending only on the mixing conductance.  The
results for the Boltzmann equation, for which we do not have analytic
results, do depend on the details of the ferromagnetic layer as seen
in Fig.~\ref{fig:analytic} for different values of the mean free path
in the ferromagnet.  When the mean free path is long so that the
conductivity in the ferromagnetic layer is much greater than that in
the non-magnetic layer, the current near the interface in the
non-magnet is increased (see Fig.~\ref{fig:dist}) giving a greater
spin Hall current.  Another difference between the approaches is that
the only length scale for variation in the drift-diffusion approach is
the spin-diffusion length.  There are many more length scales in the
Boltzmann equation approach (see Table~\ref{tab:params}) and these
turn out to play a 
non-negligible role when $\ell_{\rm N}^{\rm sf}{\rm
  Re}[g^{\uparrow\downarrow}]/\sigma_{\rm N}$ is not large.  The
deviation between the results of the Boltzmann equation calculations
and those found from the drift-diffusion equation should provide a
note of caution for the extraction of physical parameters, like the
spin Hall angle, from comparisons between experiment and the
drift-diffusion equation.

We conclude that without additional spin-orbit coupling at the
interface between the two materials, three-dimensional transport
models predict a torque that is predominantly damping-like but has a
minor field-like contribution.  In the parameter range we have
studied, the torque is always well described by the combination of
these two forms.  The drift-diffusion approach qualitatively captures
the physics but can quantitatively fail in physically relevant
parameter regimes.

\section{Inclusion of interfacial Rashba coupling}
\label{sec:rashba}

The results in 
Ref.~\onlinecite{RashbaDistorted} 
show that there is an interfacial
region with significant spin orbit coupling and exchange splitting.
In this section, we model that overlap region by adding a Rashba term
to the energy at the interface, see Eq.~(\ref{eq:intpot}).  We find
that this additional term primarily leads to a field-like torque and
that as long as it is not too strong it does not significantly modify
the torques due to the spin Hall effect.

Previously, this region has been treated by two-dimensional
calculations in which the electronic structure is modified by the
competition between the Rashba interaction and the exchange
interaction.\cite{Obata:2008,Manchon:2008,Manchon:2009,MatosAbiague:2009}
Typically, the Rashba interaction and the exchange
interaction are taken to be very different in magnitude so that the
Fermi surfaces remain essentially circular.  However, the spin
eigendirections on the Fermi surfaces are modified so that the
non-equilibrium occupation due to a current flow give rise to a net
spin accumulation that is not aligned with the magnetization.
This net transverse spin density generates a field-like exchange
torque on the magnetization.

In the Boltzmann equation approach that we use, the Rashba interaction
modifies the boundary conditions for the distribution functions
at the interface.  The resulting torque is very similar to what is
found in the two-dimensional calculations.
Depending on the details of the parameters, the transmission
probability is, on average, either greater or lesser for spins aligned
with $\hat{\bf z}\times{\bf j}$ than for those in other directions.
The spin density at the interface is 
determined by the transmission probabilities and the incident fluxes.
The bias in the transmission probabilities favors a net spin
polarization aligned with $\hat{\bf z}\times{\bf j}$, very similar to
the behavior found in the two-dimensional treatments.  Then, through
Eq.~(\ref{eq:int_torque}), there is a field-like torque.

The effect of introducing the Rashba-term on the current distribution
is shown in Fig.~\ref{fig:dist}.  Unfortunately, the spin densities at
the interface are obscured by the approximations of the Boltzmann
equation.  In this approach, we assume that electrons on different
parts of the Fermi surface are incoherent with each other.  However,
the matching 
conditions for the distribution functions across the interface are
found through coherent scattering calculations.  Once the scattering
states are used to construct the matching conditions, the coherence
between the incoming and outgoing states is neglected. As a result,
for each electron the spin density at the interface is
equal to the incident amplitude times either the transmission
probability $|T|^2$ or $|1+R|^2$, which are the same because the wave
function is continuous across the interface.  However, immediately
outside the interface, the incident and reflected states are no longer
treated as coherent, on one side of the interface the spin density is
proportional to $|T|^2$ and on the other $1+|R|^2$.  Since there are
electrons incident from both sides, the spin density at the interface
is not equal to the spin density on either side.

In Fig.~\ref{fig:dist}, we have used a strength of the Rashba
interaction that gives a torque that is comparable to that found from
the spin Hall effect.  However, this interaction only has a small
effect on the spin currents and accumulations.  We find that the spin
currents for systems with both the spin Hall effect and the
interfacial Rashba interaction can be simply and
accurately approximated as follows.  The currents in the plane, panels
a, d, and g in Fig.~\ref{fig:dist} are essentially the same in all
three systems.  For system with both the spin Hall effect and the
Rashba interaction, the transverse spin currents perpendicular to the
plane, panel e in Fig.~\ref{fig:dist}, are a sum of the transverse
spin currents found in systems with one effect or the other, panels b
and h.  The same holds true for the resulting torques, the torques for
systems with both effects are approximately the sum of the torques
found in the systems with one effect or the other.

\begin{figure}
\includegraphics[width=0.9\columnwidth]{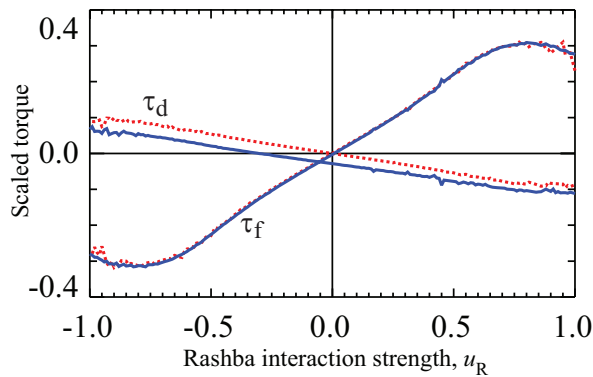}
\caption{(color online)  Scaled torques as a function of the
  interfacial Rashba interaction.  The solid curves are calculated
  with a bulk spin Hall effect and the dotted curves without.  The
  non-magnetic layer is 6~nm thick and the ferromagnetic layer is
  4~nm.  The damping-like and field-like torques are labeled with
  $\tau_{\rm d}$ and $\tau_{\rm f}$ respectively.  The jitter is
  due to numerical instabilities.} \label{fig:rashba}
\end{figure}

The approximate independence of the torques due to the spin Hall
effect and the interfacial Rashba interaction is illustrated in
Fig.~\ref{fig:rashba} as a function of the strength of the interfacial
Rashba interaction.  There are contributions to both the field-like
and damping-like torques that increase linearly with the Rashba
interaction strength up to large values of the torques.  Eventually,
the transmission probabilities get so low, see Eq. ~(\ref{eq:tramps}),
that all electrons are reflected and the torques go to zero.
Comparing calculations done with and without the spin Hall effect
shows that the interfacial coupling has very little effect on the
torque from the spin Hall effect, particularly for small values of the
Rashba interaction strength.  Without {\it a priori} knowledge of the
parameters of the system, particularly the spin diffusion length, the
spin Hall angle, and the interfacial spin-orbit coupling, a wide
variety of combinations of damping-like and field-like torques are
possible. 

\begin{figure}
\includegraphics[width=0.8\columnwidth]{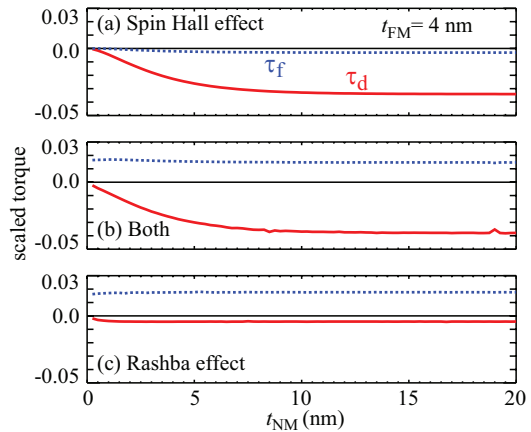}
\caption{(color online) Dimensionless torque components as a function of the
  thickness of the non-magnetic layer.  Panel (a) shows the torques in
  the absence of the Rashba contribution from the interfacial
  spin-orbit coupling, panel (c) shows the torques in the absence of
  the spin Hall effect in the non-magnet, and panel (b) shows the
  torques with both present.  In each panel, the solid lines show the
  damping-like torque and the dotted lines the field-like torque.}
\label{fig:torque_tnm}
\end{figure}

Figure~\ref{fig:torque_tnm} shows the torques as a function of the
thickness of the non-magnetic layer.  The torques due to the spin Hall
effect largely depend exponentially on the non-magnetic layer
thickness divided by the spin diffusion length. As can be seen from
the analytic solution presented in Appendix~\ref{sec:analytic}, for
very small thicknesses there are corrections such that the torque
varies as $t_{\rm NM}^2$ rather than linearly as it would if the
behavior were strictly exponential.  The exponential variation has
been used\cite{Liu:arXiv} to extract the spin diffusion length for
particular systems.  It is interesting to note that for this simple
model, the sign of the 
product $\tau_{\rm d} \tau_{\rm f}$ is opposite in the two limiting
cases, the spin  Hall effect only case [Fig.~\ref{fig:torque_tnm}(a)]
and the Rashba effect only case [Fig.~\ref{fig:torque_tnm}(c)].

\begin{figure}
\includegraphics[width=0.8\columnwidth]{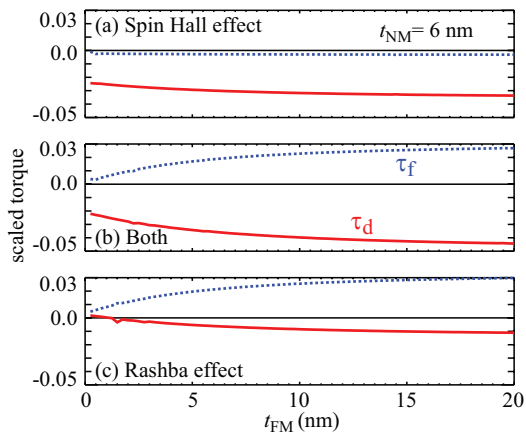}
\caption{(color online) Dimensionless torque components as a function of the
  thickness of the ferromagnetic layer.  Panel (a) shows the torques in
  the absence of the Rashba contribution from the interfacial
  spin-orbit coupling, panel (c) shows the torques in the absence of
  the spin Hall effect in the non-magnet, and panel (b) shows the
  torques with both present.  In each panel, the solid lines show the
  damping-like torque and the dotted lines the field-like torque.}
\label{fig:torque_tfm}
\end{figure}

Figure~\ref{fig:torque_tfm} shows the torque as a function of the
ferromagnetic thickness.  As expected from the analytic solution in
Appendix~\ref{sec:analytic} that is displayed in
Fig~\ref{fig:analytic}, the torque due to the spin Hall effect only
depends weakly on the thickness of the ferromagnetic layer.  The
variation is largely due to the variation in the current in the
non-magnetic layer due to the presence (and variation) of the
ferromagnetic layer.  The torque due to the
Rashba effect depends more strongly on the ferromagnetic thickness,
but not nearly so strongly as is seen in experiment where changing the
thickness by a single atomic layer can have a profound effect on the
torque.

This discrepancy suggests that the thickness dependence seen in
experiment is likely due to physics beyond the scope of the model
presented here.  One possibility is that the strength of the Rashba
interaction could depend sensitively on the thickness of the
ferromagnetic layer.  The sensitivity could arise from changes in the
electronic structure of the interface or even changes in the structure
of the interface.  Since the lattice mismatch is so large, it is
conceivable that the structure evolves rapidly as the layer is made
thicker.

\section{Summary}
\label{sec:summary}

In this article, we have developed semiclassical models for electron
and spin transport in bilayer nanowires with a ferromagnetic layer
and a non-magnetic layer with strong spin-orbit coupling.  We use a
Boltzmann equation approach, based on a simplified electronic
structure and also a simpler drift-diffusion model.  The drift-diffusion
framework qualitatively describes the physics of these systems and
provides a useful language to discuss their behavior.  However, it
quantitatively disagrees with the Boltzmann equation to which
it is an approximation.

The differences between the results found from the Boltzmann equation
and those from the drift-diffusion calculation arise for a couple of
reasons.  One reason is related to the failure of the drift-diffusion
calculation 
in other cases of in-plane transport.  While currents consist
of electrons moving in all directions, in the drift-diffusion
approximation, that motion in all (three-dimensional) directions is
averaged, leaving a single 
direction for the current.  A consequence of this averaging is that
the model misses the injection of current and spin current moving
parallel to the interface from one layer to the other.
In the case of CIP GMR, the lack of injected spin currents flowing
from one layer to the other eliminates any CIP GMR.  In the present case, the
approximation misses the injection 
of parallel current across the interfaces.  The injection (or
reduction) of the 
current flow in the plane of the interface can change the resulting
torque by a factor of two or more.
Another source of disagreement between the calculations is that the
distribution function has a different form
than that assumed when
formulating the boundary conditions in the magnetoelectronic circuit
theory.  The differences between the assumed distribution function and
that near the interface in the Boltzmann equation gives quantitative
differences between the two.

These models provide a framework that naturally includes both the
torques due to the bulk spin Hall effect and the spin transfer torque
and the torques due to the interfacial spin-orbit coupling.  These two
torques are the current induced torques that arise from spin-orbit
coupling and which are independent of the gradient of the
magnetization.  The models we treat are qualitatively similar to
previous models for the spin-Hall-induced torque but differ
substantially from the models used to describe the Rashba torque.
Those latter models are based on a two dimensional treatment of the transport
that gives rise to a current induced spin accumulation and a
predominantly field-like torque.  The Boltzmann equation approach
includes the interfacial spin-orbit coupling in the boundary
conditions of a three dimensional transport calculation.
Nevertheless, this approach gives very similar results to the
two-dimensional calculations.  The interfacial spin-orbit coupling
gives rise to a torque that is predominantly field-like.  Depending on
the specific parameters appropriate for a particular system, either
the field-like torque or the damping-like torque may dominate.

While the experimental situation is still controversial, there is
experimental evidence that both a damping-like torque and in some
systems a field-like torque play an important role in the dynamics.
The model developed here through the Boltzmann equation captures the
physics for both.  Unfortunately, it is difficult to make the model
predictive rather than explanatory.  While many of the transport
parameters are known for thick films, they are likely to change
significantly in thin films.  In fact, many vary with varying
thicknesses of the films.  This model does capture the variation with
the thickness of the non-magnetic layer, but does not describe the
rapid variation with ferromagnetic film thickness found in some
systems.  This behavior, coupled with the variation of behavior with
the order of growth, suggests to us that structural aspects of the
samples vary with thickness or growth order.  Examples of process that
might contribute to this variation include, interdiffusion, strain
relief, or grain size.

\acknowledgments

Prof.~K.-J.~Lee acknowledges support under the Cooperative Research
Agreement between the University of Maryland and the National
Institute of Standards and Technology Center for Nanoscale Science and
Technology, Award 70NANB10H193, through the University of Maryland.

\appendix

\section{Boltzmann Equation}
\label{app:be_scatt}

The Boltzmann equation is a semiclassical approach based on the
approximation that in some small but not too small region of space it
is possible to define electron wave packets that have both a well
defined momentum and a well defined position.  It is related to a
density matrix approach which neglects all of the coherence between
states with different wave vectors.  The basic quantity of interest is
the distribution function, $f({\bf k},{\bf r})$, which is the
probability to find an electron with wave vector ${\bf k}$ at position
${\bf r}$.

The straightforward generalization of the Boltzmann equation
for spin polarized systems is to have separate distribution functions
for up and down electrons, $f^\uparrow$ and $f^\downarrow$.  This is
the approach used by Camley and Barna\'{s}\cite{Camley:1989} to model
GMR.
For systems with spin-orbit coupling, in which spins can point
in arbitrary directions, there are two
related approaches to generalizing the distribution function.  In
analogy with the density matrix,
the distribution function can be generalized to a 2$\times$2 Hermitian
matrix in spin space,
${\sf f}$.  Alternatively, the same information can be captured by
four real distribution functions
$f_\alpha$, $\alpha=0,x,y,z$ related to ${\sf f}$ by
\begin{equation}
  f_\alpha = {\rm Tr}[ \sigma_\alpha {\sf f}] ,
\end{equation}
where
\begin{eqnarray}
  \sigma_0 &=& \left( \begin{array}{cc} 1&0\\ 0&1\end{array} \right) = {\rm I}\\
  \sigma_x &=& \left( \begin{array}{cc} 0&1\\ 1&0\end{array} \right) \\
  \sigma_y &=& \left( \begin{array}{cc} 0&-i\\ i&0\end{array} \right) \\
  \sigma_z &=& \left( \begin{array}{cc} 1&0\\ 0&-1\end{array} \right) \\
  \bm\sigma &=& \left( \sigma_x , \sigma_y , \sigma_z \right) 
\end{eqnarray}

The generalized Boltzmann equation is
\begin{eqnarray}
&&
{\partial f_\alpha \over \partial t }
+ { d {\bf r} \over d t } { \partial f_\alpha \over \partial {\bf r} }
+ { d {\bf k} \over d t } { \partial f_\alpha \over \partial {\bf k} }
+ \gamma H_\beta^{\rm ex} f_\gamma \epsilon_{\alpha\beta\gamma}
\nonumber\\
&=& { d f_\alpha  \over dt }_{\rm coll}
[ f_\beta (t,{\bf r},{\bf k},n) ] ,
\end{eqnarray}
where the collision term on the right hand side depends on all four
distribution functions.
The last term on the left hand side describes spin precession in
ferromagnetic layers, where the electron spins precess in the exchange
field ${\bf H}^{\rm ex}$.
The time
derivatives of ${\bf r}$ and ${\bf k}$ are given by
\begin{eqnarray}
\frac{ d {\bf r} }{ d t } &=& {\bf v}_{{\bf k},n} \\
\hbar \frac{ d {\bf k} }{ d t } &=&  - e  {\bf E}  ,
\end{eqnarray}
where ${\bf v}_{{\bf k},n}$ is the velocity of the electron and ${\bf
  E}$ is the electric field.  For the linearized Boltzmann equation,
we make the replacement
\begin{eqnarray}
f_\alpha({\bf k}) \rightarrow
f_{\rm eq}(\epsilon({\bf k})) \delta_{\alpha,0} + g_\alpha({\bf K}) f'_{\rm eq}(\epsilon({\bf k})) ,
\end{eqnarray}
Upper case ${\bf K}$ refers to wave vectors restricted to the Fermi
surface.  $f'_{\rm eq}$ is the energy derivative of the Fermi
function, emulating a Taylor series expansion of the distribution
function around equilibrium.

\begin{widetext}

After some standard algebra, the
linearized Boltzmann equation can be cast into the form
\begin{eqnarray}
\left[{\bf v}_{\bf K} \cdot
{ \partial g_\alpha({\bf K}_i)
\over \partial {\bf r} }
- e {\bf E} \cdot
{\bf v}_{\bf K} \delta_{\alpha,0}
+ \gamma H_\beta^{\rm eff}  g_\gamma({\bf K})
\epsilon_{\alpha\beta\gamma}
\right]
&=&
- R_{\alpha,\alpha'}({\bf K}_i) g_{\alpha'}({\bf K}_i)
+ \int_{\rm FS} d\hat{\bf K}_f
  P_{\alpha,\alpha'}({\bf K}_i,{\bf K}_f)
  g_{\alpha'}({\bf K}_f) ~~~~~
\end{eqnarray}
The first term on the right hand side is the scattering out term and
the second term is the scattering in term.  The former describes
collision processes that reduce the occupancy of a state and the latter those
that increase it.

In the ferromagnet, where we neglect spin-orbit
coupling, the scattering is diagonal in a coordinate system aligned with
the magnetization.  For the magnetization
pointing in a general direction
$(\sin\theta\cos\phi,\sin\theta\sin\phi,\cos\theta)$, the
spin-dependent scattering
matrix for the scattering out terms is
\begin{eqnarray}
  R_{\alpha,\alpha'}&=& U^{\rm T}_{\alpha,\beta} R^{\rm
    diag}_\beta\delta_{\beta,\beta'} U_{\beta',\alpha'} .
\end{eqnarray}
The diagonal scattering matrix is
\begin{eqnarray}
  R^{\rm diag} =
  (R^{\uparrow},\overline{R},\overline{R},R^{\downarrow}) ,
\end{eqnarray}
in terms of the majority and minority scattering rate
$R^{\uparrow}=1/\tau^\uparrow$ and $R^{\downarrow}=1/\tau^\downarrow$,
and the transverse scattering rate is taken to be the geometric
mean of the spin-dependent scattering 
$\overline{R}=1/\sqrt{\tau^\uparrow\tau^\downarrow}$.
The transformation matrix is
\begin{eqnarray}
U &=&
\left(
\begin{array}{cccc}
1/\sqrt{2} & 0 & 0 & 1/\sqrt{2} \\
0 & 1 & 0 & 0 \\
0 & 0 & 1 & 0 \\
1/\sqrt{2} & 0 & 0 & -1/\sqrt{2}
  \end{array}
  \right)
\left(
\begin{array}{cccc}
1 & 0 & 0 & 0 \\
0 & \cos\theta & 0 & -\sin\theta \\
0 & 0 & 1 & 0 \\
0 & \sin\theta & 0 & \cos\theta
  \end{array}
  \right)
\left(
\begin{array}{cccc}
1 & 0 & 0 & 0 \\
0 & \cos\phi & -\sin\phi & 0 \\
0 & \sin\phi & \cos\phi & 0 \\
0 & 0 & 0 & 1
  \end{array}
  \right)
\end{eqnarray}
The first matrix transforms between the majority/minority description
on one hand and the Cartesian description that is used in
the rest of the calculations.  The second two matrices rotate the
coordinate system.  

\end{widetext}

The scattering in terms are similar.  For these terms, where
spin-orbit coupling does not play a role, $P_{\alpha,\alpha'}$ is
independent of wave vector and equal to $R_{\alpha,\alpha'}$.
There is an additional contribution from spin flip scattering of the
form $R^{\rm sf}\delta_{\alpha,\alpha'}(1-\delta_{\alpha,0})$ in terms
of the spin flip scattering rate $R^{\rm sf}=1/\tau_{\rm sf}$.  The
last factor restricts the scattering to the spin distribution
functions and not the charge function.

In the non-magnet, the spin independent scattering is included through
a term of the form $R^{\rm N}\delta_{\alpha,\alpha'}$ in terms
of the scattering rate $R^{\rm N}=1/\tau$, and spin-flip
scattering through a term of the form
$R^{\rm Nsf}\delta_{\alpha,\alpha'}(1-\delta_{\alpha,0})$ in terms
of the spin flip scattering rate $R^{\rm Nsf}=1/\tau_{\rm sf}$.

Spin-orbit scattering is more complicated than the scattering
processes described above because the scattering rates depends on the
initial and final momenta.  Engel et al.\cite{Engel:2005} give the
contribution to the collision integral as
\begin{eqnarray}
{\frac{d{\sf f}}{dt}}_{\rm sH}
&=&
  \frac{n_i \hbar k_{\rm F}}{m^*}  \sum_{{\bf k}_{\rm f}}
  \frac{d \sigma}{d\Omega} [ {\sf f}({\bf k}_i) - {\sf f}({\bf k}_f) ]
\nonumber\\
&=&
  n_i \sum_{{\bf k}_{\rm f}} \frac{\hbar k}{m^*}
  \Bigg[
  I(\varphi) [ {\sf f}({\bf k}_i) - {\sf f}({\bf k}_f) ]
\\
&&
  - I(\varphi) S(\varphi) \bm\sigma \cdot
\frac{{\bf k}_i\times{\bf k}_f}{|{\bf k}_i\times{\bf k}_f|} 
  [ f_0({\bf k}_i) + f_0({\bf k}_f) ] 
  \Bigg] ,
\nonumber
\end{eqnarray}
where $\varphi$ is the angle between ${\bf k}_i$ and ${\bf k}_f$.
This form is based on assuming that the spin-orbit scattering is weak and
only keeping quantities lowest order in the spin-flip scattering.  
We follow this approximation but also include the scattering that
gives the inverse spin Hall effect.
The complete story 
story is much more complicated because there is also scattering
between different spin channels.   
We assume that $I(\varphi)$ is
a constant and that $S(\varphi)=S|{\bf k}_i\times{\bf k}_f|$, where $S$
is now a constant.  Then, in
our notation, we have
\begin{eqnarray}
{\frac{df_\alpha}{dt}}_{\rm sH}
&=&
  n_i \sum_{{\bf k}_{f}} \frac{\hbar k}{m^*}
  \big[
  I [ f_\alpha({\bf k}_i) - f_\alpha({\bf k}_f) ]
\\
&& ~~~~~~~~~~~~
  - IS
  n_\alpha
  [ f_0({\bf k}_i) + f_0({\bf k}_f) ]
\nonumber\\
&& ~~~~~~~~~~~~
  + IS \delta_{\alpha,0}
  n_{\alpha'}
  [ f_{\alpha'}({\bf k}_i) + f_{\alpha'}({\bf k}_f) ]
  \big] \nonumber\\
&=& \sum_{{\bf k}_{f}}
  [ - IS n_\alpha f_0({\bf k}_f)
  + IS \delta_{\alpha,0}  n_{\alpha'} f_{\alpha'}({\bf k}_f) ]
\nonumber
\end{eqnarray}
where ${\bf n}={\bf k}_i\times{\bf k}_f$ or
$n_\alpha=\epsilon_{\alpha.\beta,\gamma} k_{i\beta} k_{f\gamma}$.  
The terms containing $\delta_{\alpha,0}$ are the additional terms that
give the inverse spin Hall effect.  In
the second step, we have dropped the isotropic part because it is
simply another contribution to the isotropic scattering.  We also find
that for the remaining parts, the scattering out contribution is zero
because $\sum_f n_\alpha = 0$.  Finally, we absorbed the velocity and
the impurity density
factor into the scattering rate.  Translating to the notation we have
been using gives
\begin{equation}
  P_{\alpha,\alpha'} = PS [
  k_{i\beta} k_{f\gamma} \epsilon_{\alpha,\beta,\gamma} \delta_{\alpha',0}
- k_{i\beta} k_{f\gamma} \epsilon_{\alpha',\beta,\gamma} \delta_{\alpha,0} ]
\end{equation}

Given the scattering matrices, we find the general solutions of the
Boltzmann equation in each layer using the techniques described in
Ref.~\onlinecite{Xiao:2007}.  
These are matched together at the
interface through boundary conditions based on Eq.~(\ref{eq:intpot})
and Eq.~(\ref{eq:tramps}) and subjected to diffuse or specular
boundary conditions at the out interfaces.

The default values of the parameters we use are given in
Table~\ref{tab:params}.  These have been chosen to approximately have
values appropriate for Co/Pt bilayers with vacuum on either side.
These parameters are either input parameters, calculated numerically
based on the input parameters or determined analytically from them.
In the last case, the evaluated expression is given in the table.

\begin{table}
\caption{Default parameter values.  Parameters for the ferromagnet
  (F), are chosen to be roughly those for Co as in
  Ref.~\onlinecite{Bass:1999} and those for the non-magnet (N), to be
  roughly those for Pt as in Ref.~\onlinecite{Liu:arXiv}.
  $\lambda=v_{\rm F}\tau$ is a mean free path and $\ell^{\rm sf}$ is a
spin diffusion length.  The rest of the parameters are defined in
the text.}
\begin{tabular}{lcrl}
$\lambda_{\rm N}$ & input & 2.43 & nm  \\
$\lambda^{\rm sf}_{\rm N}$ & input & 14.7 & nm  \\
$\lambda_{\rm sH} $ & input & 11.8 & nm \\
$\lambda^{\uparrow}_{\rm N}$ & input & 16.25 & nm \\
$\lambda^{\downarrow}_{\rm N}$ & input & 6.01 & nm \\
$\lambda^{\rm sf}_{\rm F}$ & input & 3280 & nm \\
$\lambda_{\rm ex}$ & input & 0.258 & nm \\
$k_{\rm F}$ & input & 16 & nm$^{-1}$ \\
$u_0$ & input & 0.42645 &  \\
$u_{\rm ex}$ & input & 0.20055 &  \\
$u_{\rm R}$ & input & 0.04 &  \\
$\theta_{\rm SH}$ & computed & -0.059 & \\
$\ell^{\rm sf}_{\rm N}$ & computed & 2.57 & nm \\
$\ell^{\rm sf}_{\rm F}$& computed  & 69.3  & nm \\
$\sigma_{\rm N} $ & $\frac{e^2}{h}\frac{2\lambda_{\rm N}}{3\pi^2}
  \pi k^2_{\rm F} $ & 0.005 & nm$^{-1}\Omega^{-1}$ \\
$\rho_{\rm N}$ & $1/\sigma_{\rm N}$ & 20&  $\mu\Omega$cm \\
$\sigma_{\rm F} $ & $\frac{e^2}{h}\frac{\lambda^\uparrow_{\rm
      F}+\lambda^\downarrow_{\rm F}}{3\pi^2}
  \pi k^2_{\rm F} $& 0.02 & nm$^{-1}\Omega^{-1}$ \\
$\rho_{\rm F}$ & $1/\sigma_{\rm F}$ & 5&  $\mu\Omega$cm \\
$G^\uparrow$ & evaluated &6.66$\times 10^{14}$ & $\Omega^{-1}{\rm m}^{-2}$\\
$G^\downarrow$ & evaluated &3.96$\times 10^{14}$ & $\Omega^{-1}{\rm m}^{-2}$\\
Re$[G^{\uparrow\downarrow}]$ & evaluated & 5.94$\times 10^{14}$ &
$\Omega^{-1}{\rm m}^{-2}$ \\
Im$[G^{\uparrow\downarrow}]$ & evaluated &0.86$\times 10^{14}$
 & $\Omega^{-1}{\rm m}^{-2}$ \\
\end{tabular}
\label{tab:params}
\end{table}

\section{Analytical expression based on circuit theory}
\label{sec:analytic}

In this Section, we present an analytical expression of spin torque
caused by the spin Hall effect (only) in NM$|$FM bilayer structures
where NM has strong spin-orbit coupling and is thus subject to the
spin Hall effect. The derivation is based on the drift-diffusion
model~\cite{Brataas:2005} with the boundary condition of the
magnetoelectronic circuit theory.\cite{Brataas:2000}  This result
does not account for Rashba-like interactions at the interface but
provides a closed-form result in their absence.

The bulk transport is determined by
Eqs.(\ref{eq:DD_FMc}-\ref{eq:torque_density}) and the current across
the interface is calculated using magnetoelectronic circuit
theory.\cite{Brataas:2000,Brataas:2000b}  The charge and spin currents
satisfy the
boundary conditions at the NM$|$FM boundary ($z=0$) given by
\begin{eqnarray}
j_z &=& (G^{\uparrow}+G^{\downarrow})
{\Delta \mu} - (G^{\uparrow}-G^{\downarrow}) {\Delta \boldsymbol{\mu}_s } \cdot \hat{\bf
  M} \nonumber
\\
\hat{\bf z}\cdot{\bf Q} &=& {{\rm Re} [G^{\uparrow \downarrow}]}  (2 \Delta
\boldsymbol{\mu}_s \times \hat{\bf M} \times \hat{\bf M}) -{{\rm Im}
  [G^{\uparrow \downarrow} ]} (2 \Delta \boldsymbol{\mu}_s \times
\hat{\bf M}) \nonumber
\\
&&- (G^{\uparrow}+G^{\downarrow}) {\Delta
  \boldsymbol{\mu}_s}   +(G^{\uparrow}-G^{\downarrow}) {\Delta
  \mu} \hat{\bf M} \label{Eq:bc},
\end{eqnarray}
where $G^{\uparrow}$ and $G^{\downarrow}$ are
interface conductances for majority and minority spins, aligned
antiparallel and parallel 
to $\hat{\bf M}$ respectively, $G^{\uparrow \downarrow}$ is the mixing
conductance, $\Delta \mu = \mu(z=+0)-\mu(z=-0)$ is the chemical potential
drop over the interface, and $\Delta\bm{\mu}_{\rm s}$ is the spin
chemical potential drop across the interface.

By solving the bulk
equations, Eqs.(\ref{eq:DD_FMc}-\ref{eq:torque_density}), the boundary
conditions at the interface between the materials, Eq. (\ref{Eq:bc}),
with the additional boundary conditions, $j_z = 0$ at $z=0$, and ${\bf
  j}_s = 0$ at $z=+t_{\rm F}$ and $z=-t_{\rm N}$, where $t_{\rm F}$
and $t_{\rm N}$ are the thicknesses of FM and NM, respectively, one
obtains coefficients for two vector components of spin torque as in
Eq.~(\ref{eq:torqueform})
\begin{widetext}
\begin{eqnarray}
\tau_{\rm d} =  \theta_{\rm SH}
2(\kappa_{\rm N}-1)^2 l_{\rm sf}^N
\frac{2{\rm Im} [G^{\uparrow \downarrow}]^2
  (1+\kappa_{\rm N}^2) l_{\rm sf}^N + {\rm Re} [G^{\uparrow \downarrow}]
  \left[2(1+\kappa_{\rm N}^2) l_{\rm sf}^N {\rm Re} [G^{\uparrow
    \downarrow}]+(1-\kappa_{\rm N}^2)\sigma_{\rm N} \right] }
{\left[2{\rm Im} [G^{\uparrow \downarrow}] (1+\kappa_{\rm N}^2)
  l_{\rm sf}^N\right]^2+\left[2(1+\kappa_{\rm N}^2) l_{\rm sf}^N {\rm Re} [G^{\uparrow
    \downarrow}]+(1-\kappa_{\rm N}^2)\sigma_{\rm N}\right]^2}
\label{eq:analytic_td}
\\
\tau_{\rm f} = \theta_{\rm SH}
2(\kappa_{\rm N}-1)^2 l_{\rm sf}^N
\frac{{\rm Im} [G^{\uparrow \downarrow}] (1-\kappa_{\rm N}^2)
  \sigma_{\rm N} }
{\left[2{\rm Im} [G^{\uparrow \downarrow}] (1+\kappa_{\rm N}^2)
  l_{\rm sf}^N\right]^2+\left[2(1+\kappa_{\rm N}^2) l_{\rm sf}^N {\rm Re} [G^{\uparrow
    \downarrow}]+(1-\kappa_{\rm N}^2)\sigma_{\rm N}\right]^2},
\label{eq:analytic_tf}
\end{eqnarray}
where $\theta_{\rm SH} = \sigma_{\rm SH} / \sigma_{\rm N}$ is the spin
Hall angle, $\sigma_{\rm N}$ is the conductivity of NM, $\kappa_{\rm
  N}=\exp(-t_{\rm N}/l_{\rm sf}^N)$, and $l_{\rm sf}^N$ is the spin
diffusion length of NM. We note that both $\tau_{\rm d}$ and $\tau_{\rm f}$
do not depend on the magnetization direction for the case with the
spin Hall effect only.  Thus for the case with the spin Hall effect
only, the angular dependence of the torque on the magnetization
direction is completely determined by the cross products in
Eq.~(\ref{eq:torqueform}).  These damping and field like torques are 
plotted in Fig.~(\ref{fig:analytic}) (dashed lines).  These results
are independent of the thickness of the ferromagnetic layer because
the boundary conditions force the transverse spin current to be zero
in the ferromagnet.  All of the dependence of the non-magnetic layer
thickness is captured in the factors of $\kappa_{\rm N}$.

The mixing conductance in Eq.~(\ref{eq:analytic_td}) and
Eq.~(\ref{eq:analytic_tf}) is evaluated by the
integral\cite{Brataas:2000} 
\begin{eqnarray}
  G^{s,-s}=\frac{e^2}{h}\int\limits_{\rm FS} \frac{d^2k}{(2\pi)^2} (1-r_s r_{-s}^*) .
\end{eqnarray}
Here, ``up'' and ``down'' spins are defined with respect to the
magnetization, so that $s=1$ corresponds to a minority electron with
its spin parallel and moment antiparallel to the magnetization.  FS
refers to integrating over the Fermi surface.  The mixing conductance
becomes 
\begin{eqnarray}
  G^{\uparrow\downarrow}=\frac{e^2}{h}\int\limits_{\rm FS} \frac{d^2k}{(2\pi)^2}
  (1-r_\downarrow r_{\uparrow}^*) . 
\end{eqnarray}
On the right hand side, $\uparrow$ and $\downarrow$ refer to majority
and minority electrons respectively.  For the model treated here, with
a delta function sheet potential, Eq.~(\ref{eq:tramps}) gives the
reflection amplitudes, so the mixing conductance becomes
\begin{eqnarray}
  G^{\uparrow\downarrow}&=&\frac{e^2 k_{\rm F}^2}{2\pi h}
\int_0^1 dx x
\left(1-\frac{u^\downarrow}{ix-u^\downarrow}\frac{u^\uparrow}{-ix-u^\uparrow}
\right)
\nonumber\\
&=&\frac{e^2 k_{\rm F}^2}{2\pi h}
\Bigg\{
\frac{1}{2} 
+\frac{u^\uparrow u^\downarrow}{2(u^\uparrow + u^\downarrow)}
\left[ u^\downarrow {\rm ln}\left(
\frac{{u^\downarrow}^2}{1+{u^\downarrow}^2} \right) + 
       u^\uparrow  {\rm ln}\left( 
\frac{{u^\uparrow}^2}{1+{u^\uparrow}^2}  \right) \right]
\nonumber\\
&&
~~~~~~~~~~~~~~+i\frac{u^\uparrow u^\downarrow}{2(u^\uparrow + u^\downarrow)}
\left[
  u^\downarrow (\pi-2\tan^{-1}(u^\downarrow))
- u^\uparrow (\pi-2\tan^{-1}(u^\uparrow))
\right]
\Bigg\} .
\end{eqnarray}
Here $u^\uparrow=u_0-u_{\rm ex}$ and $u^\uparrow=u_0+u_{\rm ex}$.  If
$u^\downarrow > u^\uparrow$ as is the case here, both the real
and the imaginary parts of the mixing conductance are positive.
\end{widetext}

\end{document}